\documentclass[sigconf]{acmart} 
\usepackage{graphicx}
\usepackage{epstopdf}
\usepackage{threeparttable}
\usepackage{url}
\usepackage{makecell}
\usepackage{epsfig,graphicx,subfigure}
\usepackage{wrapfig}
\usepackage{tcolorbox}
\usepackage{amsmath,amsfonts}
\usepackage{algorithmic}
\usepackage{multirow}
\usepackage{array}
\usepackage{ragged2e}
\usepackage[ruled,linesnumbered,lined,boxed,commentsnumbered]{algorithm2e}
\usepackage{textcomp}
\usepackage{xcolor}
\usepackage{colortbl}
\usepackage{amsthm}
\usepackage{enumitem}
\usepackage{graphicx}
\usepackage{subcaption}
\definecolor{mygray}{gray}{0.9}
\definecolor{yellowcolor}{RGB}{191,144,0}
\AtBeginDocument{%
  \providecommand\BibTeX{{%
    \normalfont B\kern-0.5em{\scshape i\kern-0.25em b}\kern-0.8em\TeX}}}

\renewcommand\footnotetextcopyrightpermission[1]{}
\setcopyright{acmlicensed}
\copyrightyear{2018}
\acmYear{2018}
\acmDOI{XXXXXXX.XXXXXXX}

\acmConference[Conference '24]{}{ }{ }
\acmISBN{978-1-4503-XXXX-X/18/06}

\newcommand{\graphcoder}{GraphCoder\xspace}
\newtheorem{definition}{Definition}

\begin{document}

\title{\graphcoder: Enhancing Repository-Level Code Completion via Code Context Graph-based Retrieval and Language Model}

\author{Wei Liu}
\authornote{The first three authors contributed equally to this work.}
\email{weiliu@stu.pku.edu.cn}
\orcid{0009-0005-4603-4227}
\affiliation{
 \institution{Key Lab of High Confidence Software Technologies (PKU), MoE, China}
 \country{}
}
\affiliation{%
 \institution{School of Computer Science, PKU
}
 \city{Beijing}
 \country{China}
}

\author{Ailun Yu}\authornotemark[1]
\email{yuailun@pku.edu.cn}
\orcid{0009-0004-4707-4418}
\affiliation{
 \institution{Key Lab of High Confidence Software Technologies (PKU),  MoE, China}
 \country{}
}
\affiliation{%
 \institution{School of Computer Science, PKU
}
 \city{Beijing}
 \country{China}
}

\author{Daoguang Zan}\authornotemark[1]
\email{daoguang@iscas.ac.cn}
\orcid{0009-0009-4269-8543}
\affiliation{%
  \institution{Institute of Software, Chinese Academy of Sciences}
  \city{Beijing}
  \country{China}
}

\author{Bo Shen}
\email{shenbo21@huawei.com}
\orcid{0000-0002-0825-8001}
\affiliation{%
 \institution{Huawei Cloud Computing Technologies Co., Ltd.}
 \city{Beijing}
 \country{China}}

\author{Wei Zhang}
\authornote{Corresponding authors.}
\orcid{0000-0003-1543-0196}
\email{zhangw.sei@pku.edu.cn}
\affiliation{
 \institution{Key Lab of High Confidence Software Technologies (PKU),  MoE, China}
 \country{}
}
\affiliation{%
 \institution{School of Computer Science, PKU
}
 \city{Beijing}
 \country{China}
}

\author{Haiyan Zhao}
\orcid{0000-0002-3600-8923}
\email{zhhy.sei@pku.edu.cn}
\affiliation{
 \institution{Key Lab of High Confidence Software Technologies (PKU),  MoE, China}
 \country{}
}
\affiliation{%
 \institution{School of Computer Science, PKU
}
 \city{Beijing}
 \country{China}
}

\author{Zhi Jin}\authornotemark[2]
\orcid{0000-0003-1087-226X}
\email{zhijin@pku.edu.cn}
\affiliation{
 \institution{Key Lab of High Confidence Software Technologies (PKU),  MoE, China}
 \country{}
}
\affiliation{%
 \institution{School of Computer Science, PKU
}
 \city{Beijing}
 \country{China}
}

\author{Qianxiang Wang}
\orcid{0000-0002-1322-2476}
\email{wangqianxiang@huawei.com}
\affiliation{%
 \institution{Huawei Cloud Computing Technologies Co., Ltd.}
 \city{Beijing}
 \country{China}}

\renewcommand{\shortauthors}{Wei Liu, Ailun Yu, and Daoguang Zan, et al.}

\begin{abstract}
The performance of repository-level code completion depends upon the effective leverage of both \emph{general} and \emph{repository-specific} knowledge.
Despite the impressive capability of code LLMs in general code completion tasks, they often exhibit less satisfactory performance on repository-level completion due to the lack of repository-specific knowledge in these LLMs. 
To address this problem, we propose \graphcoder, a retrieval-augmented code completion framework that leverages LLMs' general code knowledge and the repository-specific knowledge via a \emph{graph-based} \emph{retrieval-generation} process. 
In particular, \graphcoder captures the context of completion target more accurately through \emph{code context graph} (CCG) that consists of control-flow, data- and control-dependence between code statements, a more structured way to capture the completion target context than the sequence-based context used in existing retrieval-augmented approaches;
based on CCG, \graphcoder further employs a \emph{coarse-to-fine} retrieval process to locate context-similar code snippets with the completion target from the current repository.
Experimental results demonstrate both the effectiveness and efficiency of \graphcoder: Compared to baseline retrieval-augmented methods, \graphcoder achieves higher exact match (EM) on average, with increases of $+6.06$ in code match and $+6.23$ in identifier match, while using less time and space.
\end{abstract}



\maketitle

\section{Introduction}

Code Large Language Models (LLMs), such as Codex~\cite{chen2021evaluating}, StarCoder~\cite{li2023starcoder} and Code Llama~\cite{roziere2023code}, have demonstrated impressive capability in general code completion tasks~\cite{zheng2023survey,nl2code,zhang2023unifying}.
These transformer-based~\cite{vaswani2017attention} large language models encode and compress extensive code knowledge into billions or even trillions of parameters through training on vast code corpora.
Some of these LLMs have been deployed as auto-completion plugins (e.g., GitHub Copilot\footnote{\url{https://github.com/features/copilot}}, CodeGeeX\footnote{\url{https://codegeex.cn}}) in modern Integrated Development Environments (IDEs), and successfully streamline the real-world software development activities to a certain degree.

However, compared with their performance in general scenarios, code LLMs exhibit less satisfactory performance in repository-level code completion tasks, due to the lack of repository-specific knowledge in these LLMs~\cite{ZanCYLKGWCL22, tang2023domain, zhang2023repocoder}. 
Specifically, the repository-specific knowledge (including code style and intra-repository API usage) cannot be well learned by or even inaccessible to code LLMs during their pre-training and fine-tuning phases, particularly for those newly created, personal privately owned, or confidential business repositories. 
One superficial remedy to this knowledge-lack problem is to concatenate all the code files in the repository as the prompt to LLMs in the situation that the size of LLMs' context window is continuously growing. 
However, this kind of remedy puts too much irrelevant information into the prompt, bringing unnecessary confusion to LLMs and thus leading to degraded completion performance~\cite{yoran2023making, shi2023large}.

To mitigate the knowledge-lack problem mentioned above, several methods have been proposed following the RAG pattern of \emph{retrieval-augmented generation}~\cite{parvez2021retrieval,lu2022reacc, zhang2023repocoder}.
For each completion task, RAG first retrieves a set of context-similar code snippets from the current repository, and then injects these snippets into the prompt, with the hope of improving the generation results of code LLMs;
these retrieved snippets play the role of augmenting code LLMs with the repository-specific knowledge related to a completion task.
As a result, the effectiveness of RAG largely depends on how to define the relevance between a code snippet and a completion task.
Most existing RAG methods follow the classical NLP style and locate a set of related code snippets of a completion task by considering sequence-based context similarity. 

In this paper, we follow the RAG pattern for repository-level code completion, but explore a more structured style to locate relevant code snippets of a completion task.
Specifically, we propose \graphcoder, a graph-based RAG code completion framework.
The key idea of \graphcoder is to capture the context of a completion task by leveraging the structural information in the source code via an artifact called \emph{code context graph (CCG)}.
In particular, a CCG is a statement-level multi-graph that consists of a set of statements as vertices, as well as three kinds of edges between statements, namely \emph{control flow}, and \emph{data}/\emph{control dependence}.
The CCG contributes to improving retrieval effectiveness from three aspects:
(1) Replacing sequence representation of code with structured representation to capture more relevant statements of the completion task;
(2) Augmenting the sequence-based similarity between the context of two statements with structure-based similarity to identify deeply matched statements of the completion target from the repository;
(3) Adopting a \emph{decay-with-distance} structural similarity to weight the different importance of context statements to the completion target.  
Experiments based on 8000 real-world repository-level code completion tasks demonstrate the effectiveness of \graphcoder. 
\graphcoder more accurately retrieves relevant code snippets with increases of +6.06 in code exact match and +6.23 in identifier exact match on average compared to RAG baseline methods while using less retrieval time and database storage space.

To summarize, our main contributions are:

\begin{itemize}[leftmargin=*]
    \item An approach \graphcoder to enhance the effectiveness of retrieval by a coarse-to-fine process, which considers both structural and lexical context, as well as the dependence distance between the completion target and the context;
    \item A graph-based representation CCG (code context graph) of source code to capture relevant long-distance context for predicting the semantics of code completion target instead of the widely adopted sequence-based one;
    \item Extensive experiments upon 5 LLMs and across 8000 code completion tasks from 20 repositories demonstrate that \graphcoder achieves higher exact match values with reduced retrieval time and overhead in database storage space.
\end{itemize}

\section{Related Work}

\emph{Repository-level Code Completion.}
The task of repository-level code completion is gaining significant attention for intelligent software development in real-world scenarios~\cite{liao2023context, ding2022cocomic, shrivastava2023repofusion, shrivastava2023repository, zhang2023repocoder}.
Recently, a growing number of large language
models (LLMs) have
been emerged and demonstrated superior performance in general code completion tasks~\cite{achiam2023gpt, lozhkov2024starcoder, roziere2023code, deepseek-coder}.
However, they demonstrate limited performance on repository-level code completion tasks due to a lack of knowledge~\cite{ZanCYLKGWCL22, tang2023domain, zhang2023repocoder, liu2023repobench, ding2024crosscodeeval}.
To address this issue, existing methods inject repository-level knowledge into LLMs either by fine-tuning them~\cite{ding2022cocomic, shrivastava2023repofusion} or by directly employing pre-trained models~\cite{tang2023domain, khandelwal2019generalization, lu2022reacc, zhang2023repocoder, tan2024prompt}.
Representative fine-tuning methods, such as CoCoMIC~\cite{ding2022cocomic} and RepoFusion~\cite{shrivastava2023repofusion}, train the language model using both in-file and relevant cross-file contexts to inject knowledge into LLMs. 
However, challenges persist due to the infeasibility of applying these methods to closed-source LLMs and the dynamic nature of repository-level features driven by continuous project development. 
To mitigate this problem, a series of methods that directly utilize pre-trained models have been proposed~\cite{tang2023domain, khandelwal2019generalization, lu2022reacc, zhang2023repocoder}.
Khandelwal et al.~\cite{khandelwal2019generalization} and Tang et al.~\cite{tang2023domain} propose a post-processing framework that adjusts the probability for the next token output by LMs with repository-level token frequency.
Nevertheless, these methods are sensitive to manually selected interpolated weights.
With the emergence of code LLMs demonstrating remarkable code comprehension capabilities, several approaches~\cite{lu2022reacc, zhang2023repocoder, liao2023context, tan2024prompt} have adopted a pre-processing strategy, that retrieves relevant snippets and adds them into LLMs' prompt. 
Additionally, benchmarks like RepoEval~\cite{zhang2023repocoder}, RepoBench~\cite{liu2023repobench}, and CrossCodeEval~\cite{ding2024crosscodeeval} have been introduced to advance the study in this field by systematically evaluating the performance of repository-level completion methods.

\emph{Retrieval-augmented Code Completion.}
Retrieval-augmented code completion is a technique that aims to integrate domain-specific knowledge into LLMs.
This technique typically first extracts the context of the completion target, then retrieves relevant code snippets, and finally concatenates the retrieved code snippets with the original context to guide the generation of LLM.
To model the context of completion target, most existing methods follow the basic idea of natural language processing, which directly extracts the last few lines of the completion target as its context and then uses it for retrieval~\cite{lu2022reacc, zhang2023repocoder, khandelwal2019generalization, tang2023domain, tan2024prompt}.
However, these methods ignore the intrinsic structure underlying the code.
Inspired by this, both the CoCoMIC~\cite{ding2022cocomic} and the RepoHyper~\cite{phan2024repohyper} construct the method-level graph to facilitate the retrieval step. Nonetheless, they still overlook the statement-level structure, which is crucial for understanding the semantics of the completion context.
For the generation step, it includes two distinct modes~\cite{tan2024prompt, lewis2020retrieval}: per-token and per-output generation.
In the per-token generation, a retrieval process is initiated for each generated token, so each token is associated each a unique set of retrieval code snippets, such as the methods $k$NN-LM~\cite{khandelwal2019generalization}, $k$NM-LM~\cite{tang2023domain}, and FT2Ra~\cite{guo2024ft2ra}.
Consequently, the number of retrievals grows as the length of the generated tokens increases, leading to a significant increase in retrieval time.
Additionally, the requirement to access each stage of token generation limits the method’s compatibility with closed-source LLMs.
In the per-output generation, a single set of retrieval code snippets is used to produce a whole sequence at once~\cite{lu2022reacc, zhang2023repocoder, phan2024repohyper}, thus improving the retrieval efficiency and facilitating compatibility with closed-source LLMs.

\section{Basic Concepts}
In this section, we introduce two concepts used in \graphcoder, namely \emph{code context graph} (CCG) and \emph{CCG slicing}.
The former is employed to transform a code snippet into a structured representation (i.e., a set of statements as well as a set of structural relationships between them).
Given a statement $x$ in a CCG $G$, the latter is used to extract a $G$'s subgraph that consists of $x$ and $x$'s $h$-hop depended elements as well as relationships between them.

\begin{figure*}
    \centering
    \includegraphics[width=0.9\linewidth]{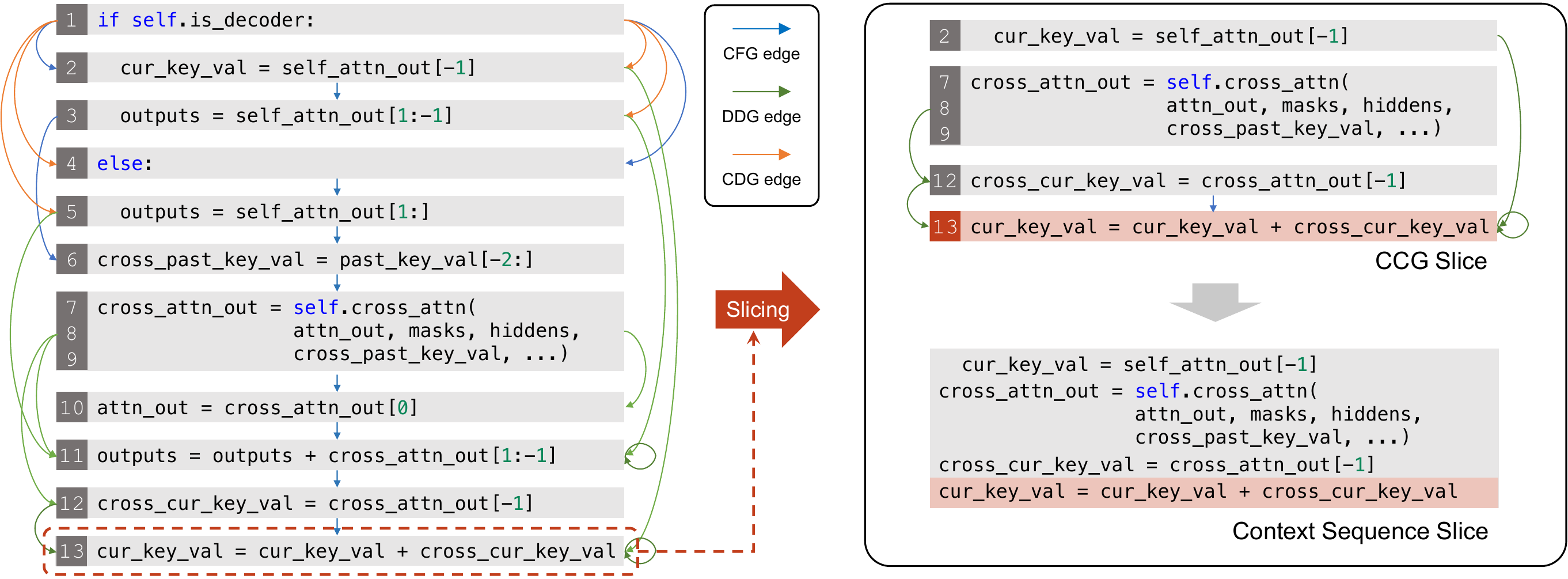}
    \caption{An example of the code context graph (CCG) and its CCG slice with statement of interest $\tilde{x}=13$.}
    \label{fig:ccg}
\end{figure*}

\subsection{Code Context Graph}

A code context graph is the superimposition of three kinds of graphs about code: control flow graph (CFG), control dependence graph (CDG), and data dependence graph (DDG). 
The latter two graphs together are commonly identified as program dependence graph~\cite{ferrante1987program}.

\begin{definition}[Code Context Graph] A code context graph $G = (X, E, T, \lambda)$ is a directed multi-graph, where
\begin{itemize}[leftmargin=*]
    \item $X = \{x_1, \cdots, x_n\}$ is the vertex set, each of which represents a code statement or a predicate;  
    \item $E = \{e_1, \cdots, e_m\}$ is the edge set; each edge is a triple $(x_i, t, x_j)$ where $x_i$, $x_j \in X$, and $t \in T$ denoting the edge type;
    \item $T = \{CF, CD, DD\}$ is the edge type set, where $CF$ denotes the \emph{control-flow} edge, $CD$ the \emph{control dependence}, and $DD$ the \emph{data dependence};
    \item $\lambda$ is a function that maps each edge in $E$ to its type in $T$, i.e., for $e = (x_i, t, x_j)$, $\lambda(e) = t$.
\end{itemize}
\end{definition}

\textbf{Control flow graphs (CFG)} 
provide a detailed representation of the order in which statements are executed~\cite{allen1970control, gold2010control, long2022multi}.
The vertices of CFG represent statements and predicates. 
The edges indicate the transitions of control between statements, including the sequential executions, jumps, and iterative loops.
The construction of a control flow graph is based on the abstract syntax tree (AST): Initially, statements and predicates are identified, and the sequential execution order is extracted from the AST. Subsequently, control transfer edges are added by analyzing conditional statements (e.g., if, for), iterative statements (e.g., for, while), and jump statements (e.g., continue, break).

\textbf{Control dependence graphs (CDG)} focus on identifying the control dependencies between statements, with edges emphasizing the direct influence of one statement on the execution of another~\cite{ferrante1987program, natour1988control, cytron1991efficiently}. Specifically, an edge exists between two statements if one directly affects the execution of another, distinguishing it from the CFG. 
Based on the CFG, CDG can be constructed by analyzing the statement reachability.

\textbf{Data dependence graphs (DDG)} reflect the dependencies arising from variable assignments and references, where edges represent that there is a variable defined in one statement is used by another~\cite{ferrante1987program, harrold1993efficient}.
The DDG can be generated through a two-step process: First, we identify the set of variables defined and used by each statement, respectively. 
Second, for a variable $v$, a DDG edge is established between two statements if there exists a CFG path from the statement defining $v$ to the statement using $v$, without intervening definitions of $v$.

\subsection{CCG Slicing} 
\begin{definition}[CCG Slice]
    Given a code context graph $G = (X, E, T, \lambda)$ and a statement of interest $\tilde{x} \in X$, the $h$-hop CCG slice of $\tilde{x}$ in $G$ with maximum $l$ statements, denoted as $G_h^l(\tilde{x})$, is defined by the output of Algorithm~\ref{alg:ccg-slicing}.
\end{definition} 

\begin{algorithm}[t]
\fontsize{8}{10}\selectfont
    \SetKwInOut{Input}{Input}
    \SetKwInOut{Output}{Output}
    \caption{CCG Slicing}
    \label{alg:ccg-slicing}
    \Input{CCG graph $G = (X, E, T, \lambda)$, statement of interest $\tilde{x} \in X$, maximum hops $h$, and maximum number of statements $l$.}
    \Output{A CCG slicing graph $G_h^l(\tilde{x})$.}
    Initialize sets $X_{CD}$, $X_{CF}$ and $X_{DD}$ as $\emptyset$\;
    Push $\tilde{x}$ into an empty queue $q$\;
    
    \While{$q$ is not empty}{
        $x \leftarrow q.pop()$\;
        \lIf{$x$ exceeds $h$ hops from $\tilde{x}$}{
            break
        }
        $X_{CF} \leftarrow X_{CF} \cup \{x\}$\;
        $X_{DD} \leftarrow X_{DD} \cup \{z \!\mid\! (z, DD, x) \in E\}$ \;
        $X_{CD} \leftarrow X_{CD} \cup \{z \!\mid\! (z, CD, x) \in E\}$ \;
        \lIf{$|X_{CF} \!\cup\! X_{CD} \!\cup\! X_{DD}| \!\geq\! l$}{
            break
        }
       \For{$z \!\in\! \{z \!\mid\! (z, CF, x) \!\in\! E, z \!\notin\! X_{CF}\}$}{
            \lIf{$z$ has not been visited by $q$}{
            $q.push(z)$
        }
       }
    }
    $G_h^l(\tilde{x}) \leftarrow G[X_{CF} \cup X_{DD} \cup X_{CD}]$\;
    
    \Return $G_h^l(\tilde{x})$
\end{algorithm}

Algorithm~\ref{alg:ccg-slicing} outlines the CCG slicing process to capture the context of a given statement $\tilde{x}$ in graph $G$. 
The key idea is to extract an induced subgraph of $G$ with vertices within $h$ hops of control-flow neighbors of $\tilde{x}$, along with the vertices they have data and control dependence on, limited to a maximum of $l$ vertices.
Starting from $\tilde{x}$ (lines 2, 3, and 5), Algorithm~\ref{alg:ccg-slicing} first updates current visited control-flow neighbors set $X_{CF}$ (line 7), and then adds its data dependence (DD) in-neighbors to $X_{DD}$ (line 8) and its control dependence(CD) in-neighbors to $X_{CD}$ (line 9).
After that, Algorithm~\ref{alg:ccg-slicing} pushes its control-flow (CF) in-neighbors to queue for the next traversing step (lines 11-13).
The final output of Algorithm~\ref{alg:ccg-slicing} is the induced subgraph of $G$ whose vertex set is $X_{CF} \cup X_{CD} \cup X_{DD}$.

Fig.~\ref{fig:ccg} provides an example of a code snippet along with its corresponding CCG and a CCG slice. The code snippet, comprising 13 lines, contains a total of 11 statements, 11 $CF$ edges, 9 $DD$ edges, and 4 $CD$ edges. Focusing on a statement of interest (line 13), its one-hop CCG slice includes all statements it has data and control dependence on (lines 2, 12, and 13), as well as its one-hop control-flow in-neighbor (line 12) and its in-neighbor's data and control dependence (lines 7-9). The context sequence slice consists of all statements in the CCG slice, ordered by line number.

\section{GraphCoder}

\subsection{Overview}
\begin{figure}
    \centering
    \includegraphics[width=0.8\linewidth]{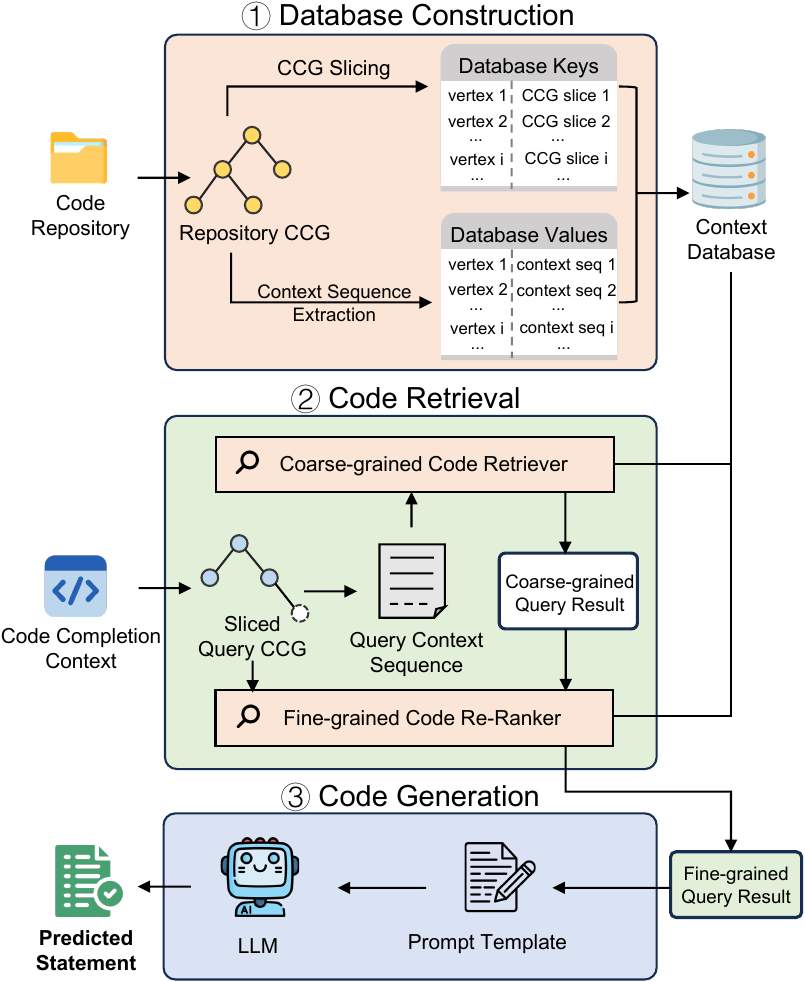}
    \vspace{-2mm}
    \caption{An illustration of \graphcoder framework.}
    \label{fig:overview}
\end{figure}

\graphcoder is a graph-based framework for repository-level code completion tasks.
In general, a code completion task aims to predict the next statement $\tilde{y}$ for a given context $X = \{x_1, x_2, \cdots, x_n\}$.
Fig.~\ref{fig:overview} gives an overview of \graphcoder's workflow.
Given a context in a code repository, \graphcoder completes the code through three steps: \emph{database construction}, \emph{code retrieval}, and \emph{code generation}.
\begin{itemize}[leftmargin=*]
    \item In the \emph{database construction} step (Section~\ref{sec:database}), \graphcoder constructs a key-value database that maps each statement's CCG slice to the statement's forward and backward $l$ lines of code.
    \item In the \emph{code retrieval} step (Section~\ref{sec:retrieval}), \graphcoder takes a code completion context as input and retrieves a set of similar code snippets through a \emph{coarse-to-fine} grained process.
    In the coarse-grained sub-process, \graphcoder filters out top-$k$ candidate code snippets based on the similarity of context sequence slice;
    in the fine-grained sub-process, the candidate snippets are re-ranked by a \emph{decay-with-distance} structural similarity measure.
    \item In the \emph{code generation} step (Section~\ref{sec:generation}), \graphcoder generates a prompt by concatenating the fine-grained query result and the code completion context, and then feeds the prompt into an LLM, waiting for the LLM to return a predicted statement $\tilde{y}$ of the code completion context.
\end{itemize}

\subsection{Database Construction}\label{sec:database}
Given a code repository, we establish a key-value database $\mathcal{D}$.
For each statement $x_i$ in the code repository, a key-value is generated and stored in $\mathcal{D}$:
the \emph{key} is $x_i$'s CCG slice $G_h^l(x_i)$, and the \emph{value} is $x_i$'s forward and backward $l$ lines of code, i.e., $\{x_{i-l/2}, \cdots, x_i, x_{i+l/2}\}$ centered around $x_i$.

\subsection{Code Retrieval}\label{sec:retrieval}
The code retrieval step takes a completion context $X$ as input, and outputs a set of code snippets, through three sub-steps: query CCG construction, coarse-grained retrieval, and fine-grained  re-ranking.

\emph{Query CCG construction.} 
\graphcoder initially extracts the sliced query CCG of the completion target. 
Specifically, it converts the given context $X$ to its CCG representation $G$. 
A dummy vertex $\tilde{y}$ is then added to $G$ to represent the statement to be predicted.
An assumption is made that there exists a control-flow edge from the last statement $x_n$ in $X$ to the statement to be predicted $\tilde{y}$.
The sliced query CCG is then obtained by slicing from $\tilde{y}$, denoted as $G_h^l(\tilde{y})$.

\paragraph{Coarse-grained retrieval.} 
Given a sliced query CCG $G_h^l(\tilde{y})$, the coarse-grained retrieval step outputs the top-$k$ most similar results in $\mathcal{D}$ based on coarse-grained similarity.
The coarse-grained similarity($CSim$) between $G_h^l(\tilde{y})$ and a key $G_h^l(x)$ in $\mathcal{D}$ is calculated as follows:
\begin{equation*} 
    CSim(G_h^l(\tilde{y}), G_h^l(x)) = sim(X_h^l(\tilde{y}), X_h^l(x)) 
\end{equation*}
where $X_h^l(\hat{y})$ and $X_h^l(x_i)$ denotes the context sequence slice based on $G_h^l(\hat{y})$ and $G_h^l(x_i)$, respectively. $sim$ denotes any similarity applicable to code sequences, including sparse retriever BM25~\cite{robertson2009probabilistic}, Jaccard index~\cite{jaccard1912distribution} based on the bag-of-words model, as well as dense retrievers like similarity of embeddings from CodeBERT~\cite{feng2020codebert} and GraphCodeBERT~\cite{guo2020graphcodebert}.

\paragraph{Fine-grained re-ranking.} In this step, \graphcoder re-ranks the coarse-grained query result based on the decay-with-distance subgraph edit distance.
The subgraph edit distance (SED) is the minimum cost of transforming one graph into a subgraph of another one through a series of edit operations~\cite{zeng2009comparing, ranjan2022greed}.
The subgraph edit operations include the deletion and the substitution of vertex or edges.
For a vertex $v$ and an edge $e$ in $G_h^l(\hat{y})$, the edit cost function $c(\cdot)$ is defined as follows:
\begin{itemize}
    \item Vertex deletion cost $c(v) = 1$;
    \item Vertex substitution cost $c(v,\! u) = 1-sim(v,\!u)$;
    \item Edge deletion cost $c(e) = 1$;
    \item Edge substitution cost $c(e, e') = \mathbf{1}_{\lambda(e) \neq \lambda(e')}$.
\end{itemize}
where $sim$ denotes any similarity measure for code sequences, and the substitution cost of the dummy vertex $\tilde{y}$ for any other vertex is assumed to be 0.

Since the subgraph edit distance problem is NP-hard~\cite{zeng2009comparing, he2006closure}, we calculate it by extending the quadratic-time greedy assignment (GA) algorithm~\cite{riesen2015approximate, riesen2015approximation} with a decay-with-distance factor.
Specifically, we first obtain an alignment $\mathcal{A}$ between the vertices in $G_h^l(\hat{y})$ and $G_h^l(x)$ by the GA algorithm~\cite{riesen2015approximate}.
Subsequently, we accumulate the edit costs as indicated by $\mathcal{A}$, as described in Algorithm~\ref{alg:sed}.
The aligned vertex pairs in $\mathcal{A}$ reflects the vertex substitution relationship between $X_h^l(\hat{y})$ and $X_h^l(x)$.
For a vertex $v$ in $G_h^l(\hat{y})$, we denote the $\mathcal{A}(v)$ as its aligned vertex in $G_h^l(x)$.
Let $X_{\mathcal{A}}$ be $\{v \!\mid\! v \in X_h^l(\hat{y}), (v, u) \in \mathcal{A}\}$, 
$E_{\mathcal{A}}$ be $\{e \!\mid\! e = (v, t, u) \in G_h^l(\tilde{y}), (\mathcal{A}(v), t', \mathcal{A}(u)) \in G_h^l(x)\}$, and $h(v, \tilde{y})$ be the number of hops from $\tilde{y}$ to $v$, the decay-with-distance SED determined by $\mathcal{A}$ is calculated in Alg.~\ref{alg:sed}.

\begin{algorithm}[t]
    \fontsize{8}{8}\selectfont
    \SetKwInOut{Input}{Input}
    \SetKwInOut{Output}{Output}
    \caption{Decay-with-distance SED}
    \label{alg:sed}
    \Input{Graphs $G_h^l(\hat{y})$ and $G_h^l(x)$ as well as a decay-with-distance factor $\gamma$.}
    \Output{Decay-with-distance SED between $G_h^l(\hat{y})$ and $G_h^l(x)$.}
    $SED \leftarrow 0$\;
    \For{$v \in X_{\mathcal{A}}$}{
        $SED \leftarrow SED + \gamma^{h(v, \tilde{y})}c(v, \mathcal{A}(v))$\;
    }
    \For{$v \in X_h^l(\hat{y}) \setminus X_{\mathcal{A}}$}{
        $SED \leftarrow SED + \gamma^{h(v, \tilde{y})}c(v)$\;
    }
    \For{$e=(v, t, u)\in E_{\mathcal{A}}$}{
        $SED \leftarrow SED + \gamma^{h(v, \tilde{y})}c(e, \mathcal{A}(e))$\;
    }
    \For{$e=(v, t, u)\in E_h^l(\hat{y}) \setminus E_{\mathcal{A}}$}{
        $SED \leftarrow SED + \gamma^{h(v, \tilde{y})}c(e)$\;
    }
    \Return $SED$\;
\end{algorithm}

\begin{figure}[!htbp]
    \centering
    \includegraphics[width=0.98\linewidth]{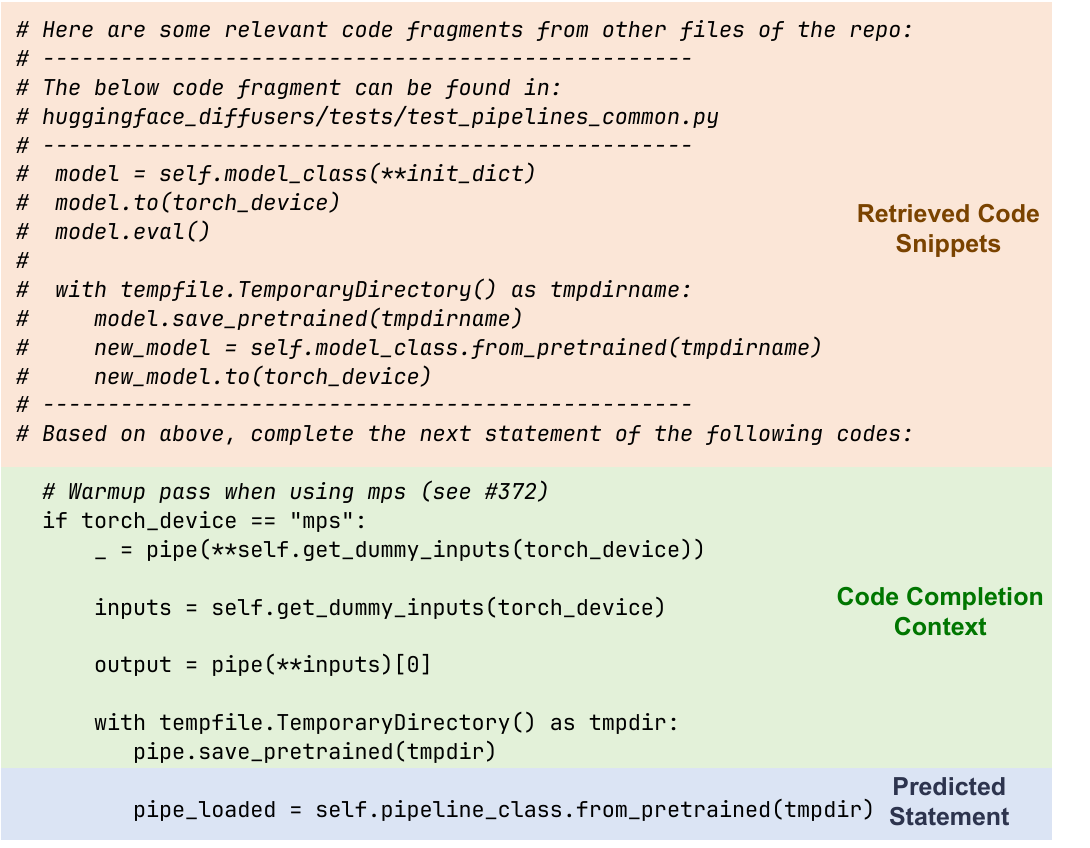}
    \vspace{-3mm}
    \caption{Prompt template used in \graphcoder.}
    \label{fig:prompt-template}
\end{figure}

\subsection{Code Generation}\label{sec:generation}
After obtaining a set of retrieved code snippets, \graphcoder employs an external LLM as a black box to generate the next statement of the given code completion context $X$. 
Following the commonly-used practice~\cite{zhang2023repocoder} of retrieval-augmented prompt formatting, we arrange the retrieval code snippets in ascending similarity order, each of which is accompanied by its original path file; then these arranged code snippets are concatenated by the code completion context $X$ as the final prompt of the LLM as shown in Fig.~\ref{fig:prompt-template}.

\section{Experimental Setup}

To evaluate the performance of \graphcoder, we have formulated the following four research questions (RQs):
\begin{itemize}[leftmargin=*]
\item \textbf{RQ1 (Effectiveness):} How does \graphcoder perform compared with other methods for repository-level code completion tasks?
\item \textbf{RQ2 (Generalizability):} How does \graphcoder perform with different base model sizes and across various repositories?
\item \textbf{RQ3 (Ablation):} How does each internal component of \graphcoder influence its performance?
\item \textbf{RQ4 (Cost):} What is the resource consumption of \graphcoder compared with other methods?
\end{itemize}

\subsection{An Updated Dataset: RepoEval-Updated}\label{sec:dataset}

The dataset RepoEval-Updated is used for repository-level code completion evaluation.
In particular, RepoEval-Updated is derived from the benchmark RepoEval~\cite{zhang2023repocoder}, which consists of a set of repository-level code completion tasks constructed from a collection of GitHub Python repositories created between 2022-01-01 and 2023-01-01.
RepoEval-Updated refreshes RepoEval by making two key changes: 
(1) It removes repositories created before March 31, 2022, and adds more recent repositories created after January 1, 2023. This update helps prevent data leakage for most existing code LLMs, whose training data was released before 2023. 
(2) Following previous work~\cite{tan2024prompt, liu2023repobench}, RepoEval-Updated adds open-source Java repositories of GitHub to include a variety of programming languages.
The details of repositories are shown in Table~\ref{tab:repos}.

Following established work~\cite{zhang2023repocoder}, we divide the repository-level code completion tasks into two categories based on their complexity, namely \emph{line-level} and \emph{API-level} tasks:
\begin{itemize}[leftmargin=*]
    \item \textbf{(Easy) Line-level tasks}: A line-level task is generated by randomly removing a code line and adhering to criteria that the target completion lines are not code comments, contain at least 5 tokens, and then encapsulating its forward code snippet as a completion task.
    \item \textbf{(Hard) API-level tasks}: An API-level task is generated in a similar way except that the removed code line includes at least one intra-repository defined API invocation.
\end{itemize}

For each task level and programming language, we randomly sample 2000 repository-level completion tasks, thus forming 8000 (2000 $\times$ 2 task levels $\times$ 2 programming languages) tasks in total.

\subsection{Evaluation Metrics}

Following the established practice~\cite{ding2024crosscodeeval, ding2022cocomic}, we evaluate the performance of RAG using the following metrics:

\begin{itemize}[leftmargin=*]
    \item \textbf{Code match}: To evaluate the level of code matching, we use two string-based metrics: exact match (EM) and edit similarity (ES). 
    The EM is a binary metric that takes the value of 1 if the predicted code equals to $y$, and 0 otherwise. The ES is a more fine-grained evaluation and is calculated as $ES=1-Lev(y, \tilde{y})/\max(|y|, |\hat{\tilde{y}}|)$, where $Lev$ represents the Levenshtein distance.
    \item \textbf{Identifier match}: We evaluate identifier matching, such as API and variable names, with two metrics: EM and F1 score.
    To calculate these two metrics, we first extract the identifiers from $\tilde{y}$ and $y$, and then directly compare their identifiers to obtain the EM and F1 scores.
\end{itemize}

{\small
\begin{table}[!t]
    \centering
    \caption{Statistics of repositories in RepoEval-Updated.}
    \vspace{-3mm}
    \scalebox{0.9}{
    \begin{threeparttable}
    \begin{tabular}{llccccc}
    \toprule
    Repo name & Created at & \#Files & Size (MB) \\
    \hline
    \href{https://github.com/devchat-ai/devchat}{devchat-ai/devchat} & 2023-04-17 & 40 & 0.5 \\
    \href{https://github.com/NVIDIA/NeMo-Aligner}{NVIDIA/NeMo-Aligner} & 2023-09-01 & 54 & 1.6 \\
    \href{https://github.com/awslabs/fortuna}{awslabs/fortuna*}  &2022-11-17 & 168 & 1.9 \\
    \href{https://github.com/microsoft/TaskWeaver}{microsoft/TaskWeaver}  & 2023-09-11 &113 & 3.0 \\
    \href{https://github.com/huggingface/diffusers}{huggingface/diffusers*}  & 2022-05-30 & 305 & 6.2 \\
    \href{https://github.com/opendilab/ACE}{opendilab/ACE*}  &2022-11-23 & 425 & 6.8 \\
    \href{https://github.com/geekan/MetaGPT}{geekan/MetaGPT}  & 2023-06-30 & 374 & 17.9 \\
    \href{https://github.com/apple/axlearn}{apple/axlearn} & 2023-02-25 & 265 & 23.8 \\
    \href{https://github.com/QingruZhang/AdaLoRA}{QingruZhang/AdaLoRA} & 2023-05-31 & 1357 & 32.6 \\
    \href{https://github.com/nerfstudio-project/nerfstudio}{nerfstudio-project/nerfstudio*}  & 2022-05-31 & 157 & 54.5 \\
    \hline
    \href{https://github.com/itlemon/chatgpt4j}{itlemon/chatgpt4j} & 2023-04-04 & 67  & 0.4 \\
    \href{https://github.com/Aelysium-Group/rusty-connector}{Aelysium-Group/rusty-connector} & 2023-02-25 & 133 & 2.6 \\
    \href{https://github.com/neoforged/NeoGradle}{neoforged/NeoGradle} & 2023-07-08 & 129 & 3.3 \\
    \href{https://github.com/mybatis-flex/mybatis-flex}{mybatis-flex/mybatis-flex} & 2023-02-27 & 487 & 8.8 \\
    \href{https://github.com/Guiqu1aixi/rocketmq}{Guiqu1aixi/rocketmq} & 2023-04-25 & 988 & 10.6 \\
    \href{https://github.com/SimonHalvdansson/Harmonic-HN}{SimonHalvdansson/Harmonic-HN} & 2023-05-23 & 51 & 16.8 \\
    \href{https://github.com/Open-DBT/open-dbt}{Open-DBT/open-dbt} & 2023-02-27 & 366 & 20.0 \\
    \href{https://github.com/QuasiStellar/custom-pixel-dungeon}{QuasiStellar/custom-pixel-dungeon} & 2023-05-12 & 1093 & 51.3 \\
    \href{https://github.com/gentics/cms-oss}{gentics/cms-oss} & 2023-05-08 & 2580 & 130.5\\
    \href{https://github.com/FloatingPoint-MC/MIN}{FloatingPoint-MC/MIN} & 2023-07-10 & 2628 & 269.5\\
    \bottomrule
    \end{tabular}
    \begin{tablenotes}
        \footnotesize
        \vspace{-0.5mm}
        \item * corresponds to the repositories of original benchmark. The former 10 code repositories are in Python, and the latter 10 are in Java. All the newly added repositories are archived on 2024-05-16. \#Files indicates the number of Python/Java files in the repository. Statistics are accurate as of May 2024.
      \end{tablenotes}
    \end{threeparttable}}
    \label{tab:repos}
\end{table}
}

\subsection{Methods for Comparison}

As \graphcoder focuses on improving retrieval results by incorporating statement-level structural information into code context instead of the widely adopted sequence-based one, we select the following four methods to evaluate its effectiveness:

\begin{itemize}[leftmargin=*]
    \item \textbf{No RAG.} This method simply feeds the code completion context into an LLM and takes the output of the LLM as the predicted next statement.
    \item \textbf{Vanilla RAG.} Given a context, this method retrieves a set of similar code snippets from a repository via a fixed-size sliding window and invokes an LLM to obtain a predicted next statement.
    \item \textbf{Shifted RAG.} This method is similar to vanilla RAG, except that it returns the code snippet in the subsequent window that is more likely to include the invocation example of target code. 
This method is also mentioned in ReAcc~\cite{lu2022reacc}.
    \item \textbf{RepoCoder}~\cite{zhang2023repocoder}. A sliding window-based method that locates the completion target through an iterative retrieval and generation process. In each iteration, RepoCoder retrieves the most similar code snippets based on the code LLMs' generation results from the last iteration.
\end{itemize}

\subsection{Implementation Details}

\subsubsection{Code Retrieval.} 
To ensure a fair comparison, we use the same measure to compute the similarity between code sequences across different methods for comparison.
Specifically, we employ a sparse bag-of-words model, known for its effectiveness in retrieving similar code snippets~\cite{lu2022reacc, zhang2023repocoder}, a model that transforms code snippets into sets of tokens and calculates similarity using the Jaccard Index~\cite{jaccard1912distribution}.
For sliding window-based methods (Vanilla RAG, Shifted RAG, and RepoCoder), we fix the window size as 20 lines and a default sliding stride of 1. 
For \graphcoder, its maximum hop $h$ is set to 5, the maximum number of statements $l$ is set to 20, and the decay-with-distance factor is set to 0.1.
To construct CCG, we first build the abstract syntax tree (AST) of a code snippet by utilizing tree-sitter\footnote{\url{https://tree-sitter.github.io/tree-sitter/}}, and then identify the statements within the code snippet and performing control-flow/dependencies analysis techniques.

\subsubsection{Code Generation.}
To avoid data leakage, we exclude in our consideration those LLMs without an explicit training data timestamp or a timestamp after 2023-01-01.
Among the remaining LLMs, we select 5 LLMs with diverse code understanding capabilities: GPT-3.5-Turbo-Instruct~\footnote{\url{https://platform.openai.com/docs/models/gpt-3-5-turbo}}, StarCoder 15B~\cite{li2023starcoder}, and CodeGen2 models (1B, 3.7B, 7B, and 16B)~\cite{nijkamp2023codegen2}.
Following established practice in code completion~\cite{zhang2023repocoder}, we fill the LLMs' context window with two parts: 
the retrieved code snippets, and the completion context. 
Each part occupies half of the context window. 
The maximum number of retrieved code snippets is 10.
The maximum number of tokens in the generated completion is set to 100. 
The temperature of LLMs is set to 0 to ensure reproducibility.

Notice that CCG is language-agnostic, \graphcoder can be migrated to other programming languages by following the same procedure. In our experiments, we focus on Python and Java as proof-of-concept languages to showcase \graphcoder's performance. All experiments are conducted on a cluster equipped with 14 Xeon Gold 6330 CPUs and NVIDIA A100-80GB GPU.

{\small
\begin{table*}[!ht]
    \centering
    \caption{Experimental results on the code completion effectiveness.}
    \vspace{-4mm}
    \scalebox{0.95}{
    \begin{threeparttable}
    \begin{tabular}{cclcccccccccccc}
    \toprule
    & & & \multicolumn{4}{c}{GPT3.5} & \multicolumn{4}{c}{StarCoder-15B} & \multicolumn{4}{c}{CodeGen2-16B} \\
    \cmidrule(lr){4-7}\cmidrule(lr){8-11}\cmidrule(lr){12-15}
    & & & \multicolumn{2}{c}{Code Match} & \multicolumn{2}{c}{Identifier Match} & \multicolumn{2}{c}{Code Match} & \multicolumn{2}{c}{Identifier Match} & \multicolumn{2}{c}{Code Match} & \multicolumn{2}{c}{Identifier Match} \\
    & & & EM & ES & EM & F1 & EM & ES & EM & F1 & EM & ES & EM & F1 \\
    \hline
    \multirow{10}{*}{Line-level} & \multirow{5}{*}{Python} & No RAG & 33.10 & 60.28 & 40.35 & 56.15 & 19.75 & 42.46 & 21.10 & 29.48 & 33.75 & 60.71 & 40.90 & 56.32 \\
    & & Vanilla RAG  & 37.90 & 58.47 & 44.40 & 54.30  & 32.80 & 53.28 & 36.75 & 42.27 & 39.00 & 61.15 & 45.90 & 56.58 \\
    & & Shifted RAG  & 45.65 & 68.49 & 51.70 & 63.62  & 18.70 & 36.34 & 22.35 & 25.55 & 35.65 & 55.54 & 41.00 & 50.80 \\
    & & RepoCoder    & 38.20 & 59.71 & 44.75 & 54.59  & 34.30 & 53.22 & \textbf{38.10} & 43.14 & 41.10 & 63.05 & 48.25 & 58.44 \\
    & & GraphCoder   & \textbf{46.60} & \textbf{69.42} & \textbf{53.80} & \textbf{65.55} & \textbf{34.50} & \textbf{54.11} & \textbf{38.10} & \textbf{43.24} & \textbf{46.65} & \textbf{69.29} & \textbf{53.55} & \textbf{65.25} \\
    \cmidrule(lr){2-15}
    & \multirow{5}{*}{Java} & No RAG &  43.20 & 76.01 & 51.60 & 68.31 & 23.75 & 46.07 & 34.80 & 37.39 & 32.80 & 69.09 & 43.90 & 59.56 \\
    & & Vanilla RAG  & 48.05 & 78.04 & 56.20 & 70.89 & 28.10 & 49.85 & 32.35 & 35.02 & 36.15 & 69.67 & 46.95 & 59.37 \\
    & & Shifted RAG  & 48.15 & 78.09 & 55.80 & 70.74 & 25.15 & 52.25 & 30.10 & 35.33 & 36.15 & 70.16 & 46.80 & 59.81 \\
    & & RepoCoder    & 48.30 & 78.16 & 56.50 & 70.76 & 30.15 & 51.73 & 34.35 & 36.80 & 37.70 & 70.86 & 48.80 & 60.72 \\
    & & GraphCoder   & \textbf{50.60} & \textbf{78.94} & \textbf{58.70} & \textbf{72.00} & \textbf{30.83} & \textbf{54.89} & \textbf{35.81} & \textbf{39.14} & \textbf{40.30} & \textbf{72.05} & \textbf{50.45} & \textbf{61.95}\\
    \hline
    \multirow{10}{*}{API-level} & \multirow{5}{*}{Python} & No RAG  & 27.75 & 56.55 & 30.90 & 54.33  & 15.05 & 39.49 & 15.35 & 26.16 & 27.70 & 56.61 & 30.35 & 53.43  \\
    &                        & Vanilla RAG  & 37.50 & 57.98 & 40.05 & 56.77  & 35.90 & 54.08 & 37.25 & 51.15 & 35.80 & 58.66 & 38.60 & 56.82  \\
    &                        & Shifted RAG  & 41.65 & 65.06 & 44.25 & 63.58  & 17.60 & 34.88 & 18.00 & 24.61 & 34.80 & 55.32 & 37.35 & 52.82  \\
    &                        & RepoCoder    & 39.40 & 59.28 & 42.10 & 57.88  & 36.70 & 58.30 & 40.15 & 55.74 & 41.00 & 63.07 & 44.00 & 61.45  \\
    &                        & GraphCoder   & \textbf{45.25} & \textbf{66.81} & \textbf{48.80} & \textbf{65.70}  & \textbf{38.90} & \textbf{60.37} & \textbf{41.59} & \textbf{56.15} & \textbf{48.75} & \textbf{69.97} & \textbf{51.75} & \textbf{69.03}  \\
    \cmidrule(lr){2-15} 
    & \multirow{5}{*}{Java} & No RAG & 37.95 & 71.91 & 40.70 & 63.79 & 23.95 & 52.43 & 24.50 & 37.01 & 24.25 & 60.88 & 26.45 & 47.80 \\
                            & & Vanilla RAG  & 54.10 & 79.27 & 57.00 & 73.59 & 46.50 & 63.03 & 46.70 & 52.09 & 52.35 & 75.89 & 54.10 & 70.32 \\
                            & & Shifted RAG  & 58.80 & 81.45 & 61.25 & 76.16 & 49.20 & 67.22 & 49.40 & 55.64 & 54.65 & 77.87 & 56.50 & 73.58 \\
                            & & RepoCoder    & 56.05 & 79.80 & 58.55 & 74.27 & 48.45 & 64.40 & 48.70 & 53.76 & 57.20 & 78.82 & 59.05 & 74.06\\
                        & & GraphCoder   & \textbf{61.57} & \textbf{82.66} & \textbf{63.72} & \textbf{77.68} & \textbf{54.90} & \textbf{69.85} & \textbf{55.00} & \textbf{59.83} & \textbf{60.15} & \textbf{80.53} & \textbf{61.55} & \textbf{76.34} \\
    \bottomrule
    \end{tabular}
    \begin{tablenotes}
        \footnotesize
        \vspace{-0.5mm}
        \item The values presented are formatted as percentages (\%). GPT3.5 refers to GPT3.5-Turbo-Instruct. The results of RepoCoder are obtained after three iterations.
      \end{tablenotes}
    \end{threeparttable}}
    \label{tab:completion-results}
\end{table*}
}

\section{Experimental Results}

\subsection{RQ1: Effectiveness}

In this subsection, we study the effectiveness of \graphcoder compared with baseline methods both quantitatively and qualitatively.

\subsubsection{Quantitative analysis of effectiveness}
Table~\ref{tab:completion-results} shows the completion results on line-level and API-level tasks across different methods.
Across all LLMs, \graphcoder outperforms other baselines for both API-level and line-level code completion tasks. 
This result demonstrates the benefits of utilizing the structural context extracted based on CCG for locating relevant code snippets to the completion target.
Compared to the vanilla RAG, \graphcoder increases the code match EM values on API-level and line-level tasks by $+4.58$ and $+7.90$ on average, respectively.
This observation emphasizes the effectiveness of \graphcoder's retrieval in repository-level code completion scenarios.
Furthermore, compared with other sliding window-based RAG methods (Vanilla RAG, Shifted RAG, and RepoCoder), \graphcoder exhibits superior performance with higher code match scores and identifier match scores. 
Notably, an observation from Table~\ref{tab:completion-results} indicates that Shifted RAG's shifting approach does not necessarily enhance No RAG completion performance.
However, shifting all retrieved code snippets without considering their content may lead to the retrieval of totally irrelevant code snippets, introducing potential confusion for LLMs.
Additionally, the effectiveness of RAG methods is more evident on harder API-level tasks, where performance is generally lower than on line-level tasks. This observation emphasizes the necessity of retrieving relevant code snippets from repositories for API-level tasks in real-world scenarios.

\subsubsection{Qualitative analysis of effectiveness.}

 \begin{figure}[!t]
    \centering
    \includegraphics[width=0.9\linewidth]{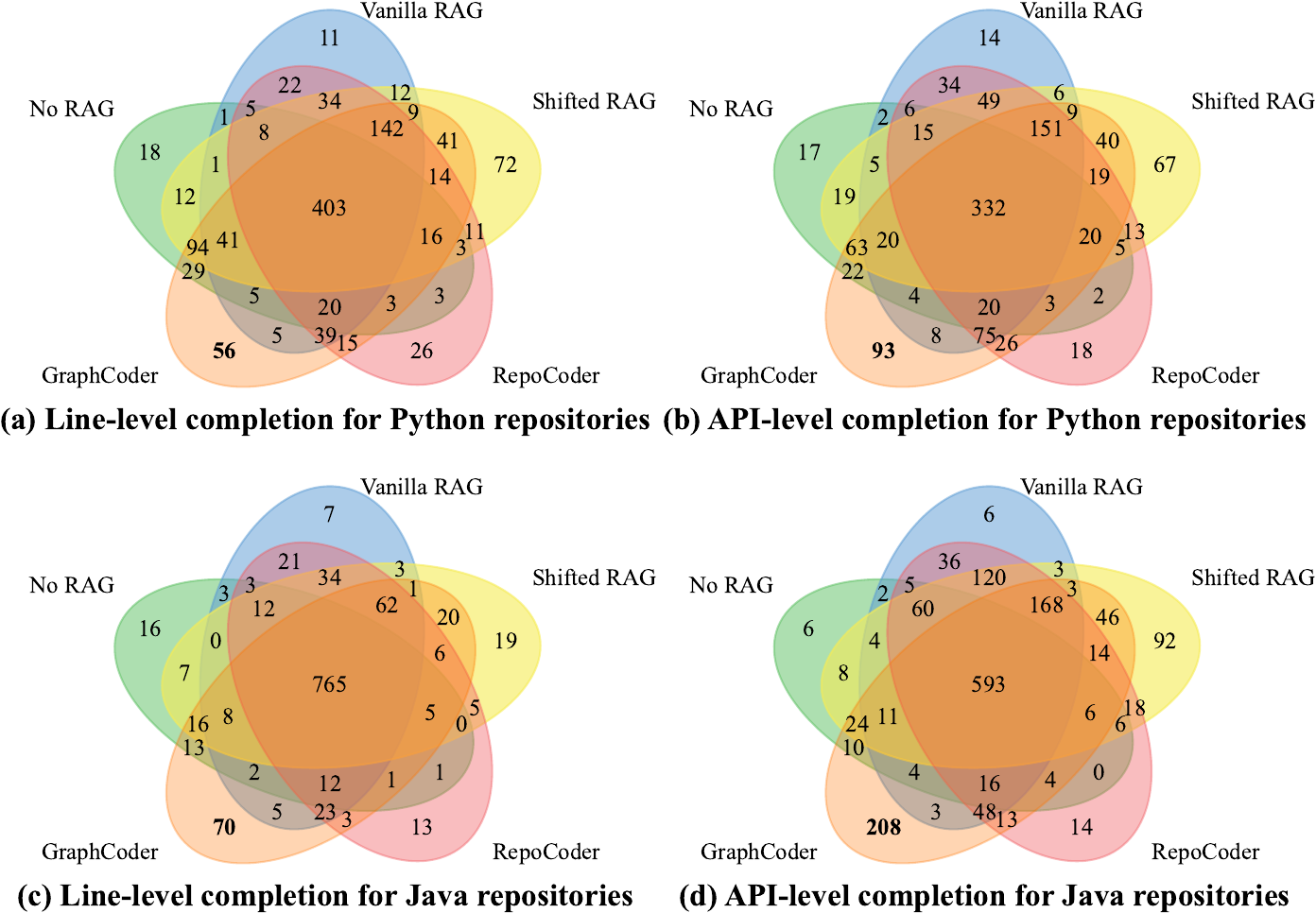}
    \vspace{-2mm}
    \caption{Venn diagram of completion results on GPT3.5-Turbo-Instruct model of different methods. It shows the number of tasks that are completed correctly.}
    \label{fig:case-analysis}
\end{figure}

\begin{figure*}
    \centering
    \includegraphics[width=0.98\linewidth]{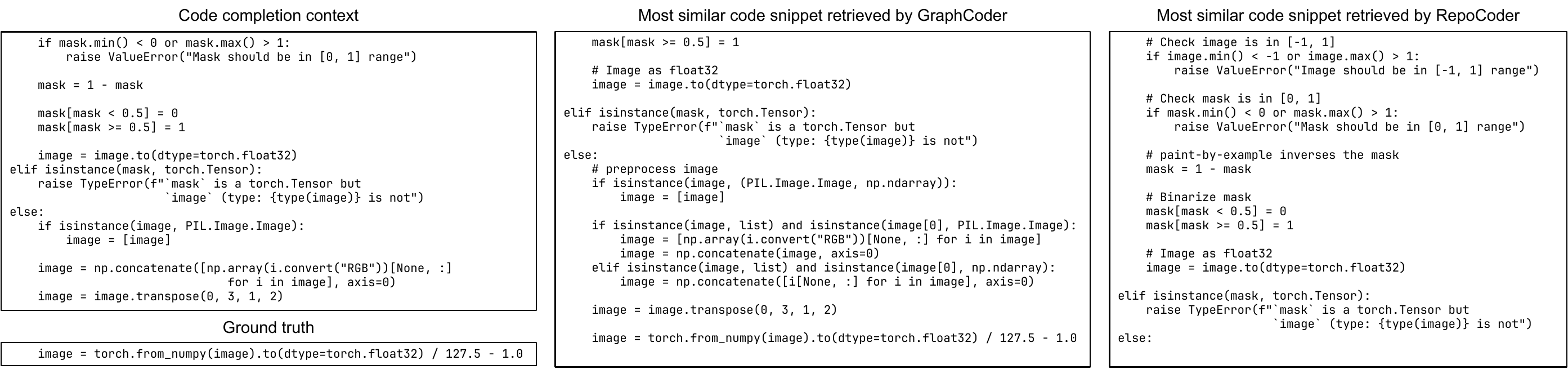}
    \vspace{-2mm}
    \caption{A qualitative example demonstrating the effectiveness of \graphcoder.}
    \label{fig:example}
\end{figure*}

To further investigate the differences among various methods, we analyze the number of tasks that they complete correctly by the Venn diagrams shown in Fig.~\ref{fig:case-analysis}.
It is observed that \graphcoder completes the highest number of tasks that none of the other methods can correctly complete on Java and API-level Python tasks.
By re-examining the experimental results, we observe that \graphcoder's superiority mainly derives from its structural alignment-based similarity (i.e., decay-with-distance subgraph edit distance) to locate structure deeply-matched code snippets. 
As shown in Fig.~\ref{fig:example}, the code snippet retrieved by \graphcoder is better aligned with the code completion context compared to RepoCoder. 
Consequently, \graphcoder's snippet provides more relevant information, enabling the LLM to correctly generate the next statement.
Additionally, there is a small proportion of tasks that are correctly completed by all RAG methods except \graphcoder. This occurs because retrieved code snippets can sometimes be misleading. Specifically, when the context of retrieved code snippets is very similar to that of the current code completion task, LLMs tend to directly copy the subsequent statement of the retrieved snippets without adapting to the slightly different completion context.

\begin{tcolorbox}[left=2pt,right=2pt,top=0pt,bottom=0pt,boxrule=0.5pt]
\textbf{Answer to RQ1: }
\textit{
On 8000 repository-level code completion tasks, 
\graphcoder demonstrates superior performance compared to baseline RAG methods, achieving a higher EM in both code match (+6.06) and identifier match (+6.23) on average.
}
\end{tcolorbox}

\subsection{RQ2: Generalizability}
In this subsection, we explore the generalizability of \graphcoder by examining its performance on base models of different sizes and on repositories with various code duplication ratios.

\subsubsection{Generalizability across base models of different sizes}
\begin{figure}[!t]
    \centering
    \includegraphics[width=0.9\linewidth]{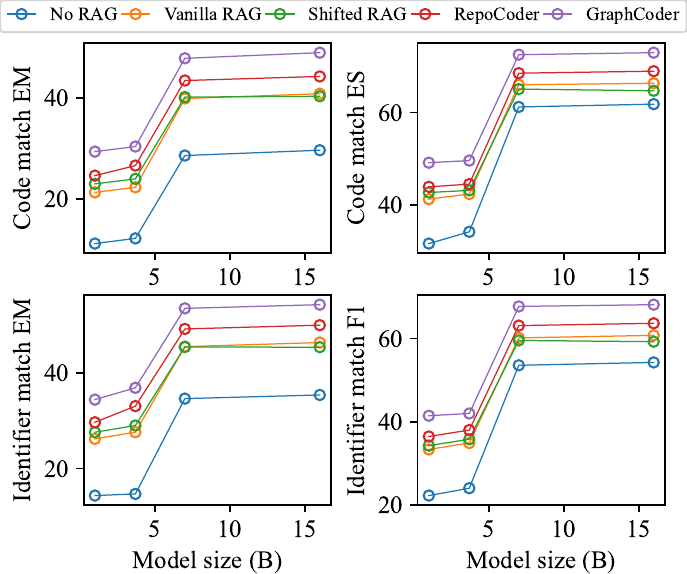}
    \vspace{-2mm}
    \caption{Performance of RAG and non-RAG methods across different base model sizes (CodeGen2 1B, 3.7B, 7B, and 16B).}
    \label{fig:scalability}
\end{figure}

Fig.~\ref{fig:scalability} shows the performance of RAG methods as the model size increases. The results are averaged across line-level and API-level tasks for both Python and Java languages. 
As model size increases, \graphcoder consistently yields the best performance among the four RAG methods, which indicates that its structural similarity-based retrieval successfully enables the model to better understand and generate code based on the retrieved code snippets with similar structures and patterns.
As model size increases, the RAG methods and No RAG tend to perform better. However, on repository-level tasks, the performance of models without RAG does not follow the scaling law~\cite{kaplan2020scaling}; the performance does not scale as a power-law with model size due to the lack of intra-repository knowledge.
Additionally, it can be observed that there is a phase transition from CodeGen2-3.7B to CodeGen2-7B; when transitioning from CodeGen2-3.7B to CodeGen2-7B, a substantial change occurs. 
In contrast, from CodeGen2-7B to CodeGen2-16B, the improvement in performance is limited. 
Therefore, considering cost-effective balance, an LLM of size 7B may be a good choice for implementing RAG-based completion methods. 

\begin{figure}
    \centering
    \includegraphics[width=0.9\linewidth]{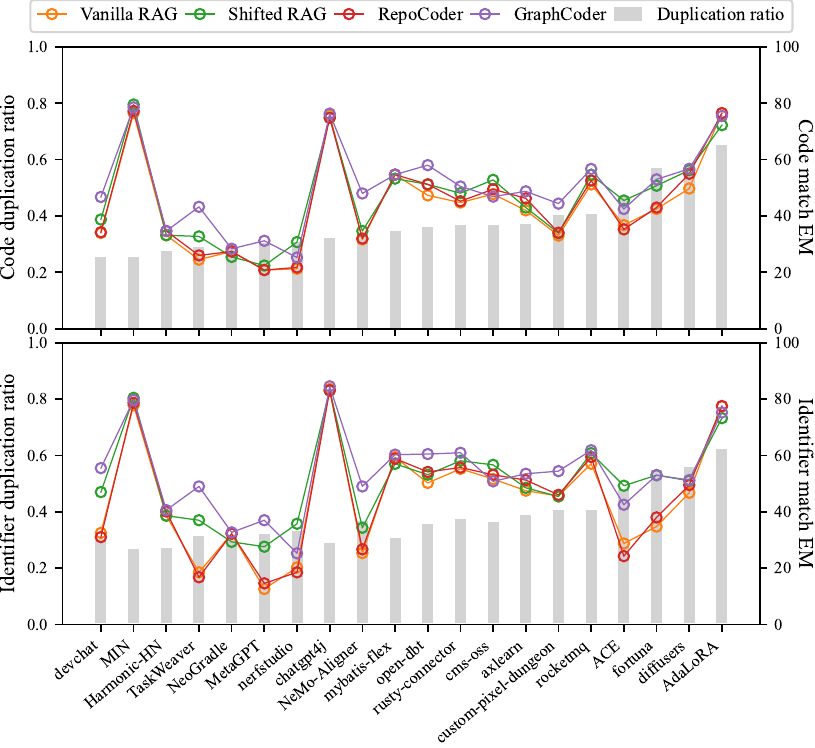}
    \vspace{-4mm}
    \caption{Correlation between the repository’s duplication ratio and the performance of RAG methods.}
    \label{fig:duplication}
\end{figure}

\subsubsection{Generalizability across repositories of various duplication ratios}
Fig.~\ref{fig:duplication} shows the correlation between the repository's duplication ratio and the performance of RAG methods.
Intuitively, it is easier to retrieve useful code snippets when a repository has a large amount of duplication.
The code duplication ratio~\cite{zhang2023repocoder} measures duplicated code lines in a repository, while the identifier duplication ratio measures lines with repeated identifiers relative to the total code lines.
The results of Fig.~\ref{fig:duplication} are based on GPT-3.5-Turbo-Instruct.
As shown in Fig.~\ref{fig:duplication}, \graphcoder outperforms other RAG methods in 15 out of 20 repositories, indicating its superior performance. 
The trend depicted by the curve suggests that repositories with lower duplication levels benefit more from \graphcoder. 
Specifically, when the code duplication ratio is below 40\%, \graphcoder surpasses the best result of baseline methods by 9.13\%, and by 5.25\% when the ratio is above 40\%. 
These results highlight \graphcoder's ability to effectively retrieve relevant code snippets compared to other baseline RAG methods, particularly in more challenging tasks with less superficial code duplication.

\begin{tcolorbox}[left=2pt,right=2pt,top=0pt,bottom=0pt,boxrule=0.5pt]
\textbf{Answer to RQ2: }
\textit{
As the model size increases, \graphcoder consistently outperforms other RAG methods and demonstrates greater effectiveness in repositories with lower repetition.
}
\end{tcolorbox}

\subsection{RQ3: Ablation Study}

In this subsection, we systematically evaluate the effects of the key components in \graphcoder. For simplification, all the experiments are conducted based on GPT3.5-Turbo-Instruct.

\subsubsection{Ablation study of components in CCG}

The three components in CCG are the control flow graph (CFG), the data dependence graph (DDG), and the control dependence graph (CDG). To study the impact of each component, we separately remove each of them and then evaluate their performance.

As shown in Table~\ref{tab:ablation-ccg}, the importance of the components follows a clear hierarchy: CFG is most critical, followed by CDG, and then DDG. 
The removal of CFG results in a significant decline in the performance, with average relative reductions of 19.82\% in code match EM and 15.64\% in identifier match EM.
This underscores the fundamental role of CFG, which is the basis for CCG construction that integrates CDG and DDG nodes through traversal.
Without CFG, the CCG slice degenerates to a 1-hop DDG and CDG local subgraph. 
The most significant drop occurs on the API-level Java tasks.
Compared to Python, the more verbose syntax of Java requires a greater need for long-distance relevant information to understand the context of code completion.
The impact of removing CDG and DDG is nearly equivalent, with a negligible difference in performance degradation, suggesting that while both are important, predicates in CDG may play a slightly more critical role in determining the semantics of their corresponding statements.

\subsubsection{Ablation study for coarse-to-fine steps.}
\begin{figure}[!t]
    \centering
    \includegraphics[width=0.82\linewidth]{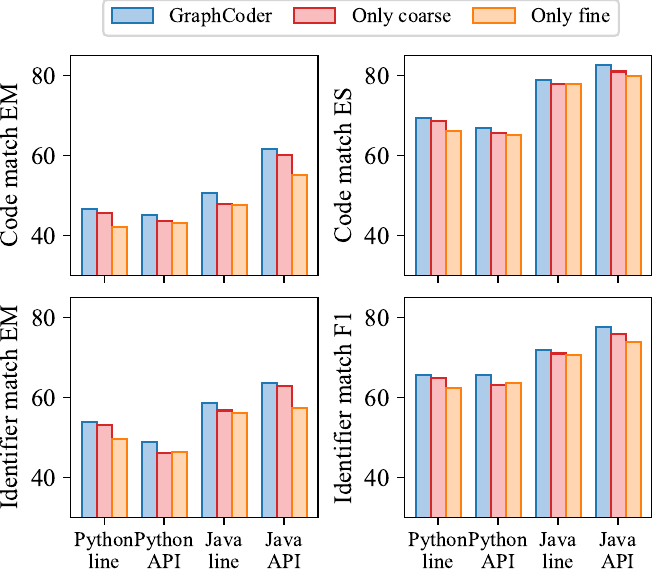}
    \vspace{-2mm}
    \caption{Ablation study for coarse-to-fine steps.}
    \vspace{-4mm}
    \label{fig:ablation-stages}
\end{figure}

Fig.~\ref{fig:ablation-stages} shows the ablation study results of the coarse-to-fine steps in \graphcoder. It evaluates their impact on \graphcoder by comparing the performance of using only the coarse-grained retrieval and only the fine-grained retrieval separately.

As seen from Fig.~\ref{fig:ablation-stages}, \graphcoder consistently exhibits better performance than its variants that employ either only coarse-grained or only fine-grained retrieval steps. 
This observation confirms the benefits of integrating both coarse-grained and fine-grained retrieval steps within \graphcoder. 
In particular, the coarse-grained retrieval step plays a more significant role than the fine-grained step.
When comparing the \graphcoder only coarse variant to the only fine-grained variant, there are notable improvements in the code match EM, ES, and identifier match EM, F1 scores of 2.34, 1.15, 2.41, and 1.08, respectively.
A particularly significant drop in performance is observed for the only fine-grained variant on the API-level Java task.
This is because, compared with the coarse-grained retrieval, the fine-grained step focuses more on localized context, while Java completion task typically relies more on long-distance relevant context
This observation is also consistent with the conclusion drawn from Table~\ref{tab:ablation-ccg}.

{\small
\begin{table}[!t]
    \centering
    \caption{Ablation study of components in CCG.}
    \vspace{-2mm}
    \scalebox{0.88}{
    \begin{tabular}{lcrcccc}
    \toprule
    & & & \multicolumn{2}{c}{Code Match} & \multicolumn{2}{c}{Identifier Match} \\
    & & & EM & ES & EM & F1 \\
    \hline
    \multirow{8}{*}{\makecell{Line\\level}} & \multirow{4}{*}{Python} & \textbf{GraphCoder} & \textbf{46.60} & \textbf{69.42} & \textbf{53.80} & \textbf{65.55} \\
    & & - C\hspace{0.1em}F\hspace{0.1em}G  & 39.15 & 62.93 & 47.10 & 53.93 \\
    & & - DDG  & 42.05 & 64.96 & 49.85 & 56.21\\
    & & - CDG  & 41.70 & 64.89 & 49.50 & 56.16\\
    \cmidrule{2-7}
    & \multirow{4}{*}{Java} & \textbf{GraphCoder} & \textbf{50.60} & \textbf{78.94} & \textbf{58.70} & \textbf{72.00} \\
    & & - C\hspace{0.1em}F\hspace{0.1em}G   & 45.07 & 77.10 & 54.18 & 69.67\\
    & & - DDG  & 47.62 & 77.66 & 56.20 & 70.27\\
    & & - CDG  & 47.56 & 77.62 & 56.09 & 70.26\\
    \hline
    \multirow{8}{*}{\makecell{API\\level}} & \multirow{4}{*}{Python} & \textbf{\graphcoder} & \textbf{45.25} & \textbf{66.81} & \textbf{48.80} & \textbf{65.70} \\
    & & - C\hspace{0.1em}F\hspace{0.1em}G   & 35.90 & 61.01 & 42.40 & 51.65\\
    & & - DDG  & 39.80 & 63.51 & 46.40 & 54.66\\
    & & - CDG  & 39.60 & 63.11 & 46.20 & 54.27\\
    \cmidrule{2-7}
    & \multirow{4}{*}{Java} & \textbf{GraphCoder} & \textbf{61.57} & \textbf{82.66} & \textbf{63.72} & \textbf{77.68} \\
    & & - C\hspace{0.1em}F\hspace{0.1em}G  & 42.06 & 73.90 & 45.06 & 66.19 \\
    & & - DDG  & 56.62 & 79.72 & 58.77 & 75.77 \\
    & & - CDG  & 56.36 & 79.61 & 58.62 & 75.69 \\
    \bottomrule
    \end{tabular}}
    \label{tab:ablation-ccg}
\end{table}
}

\subsubsection{Hyper-parameter sensitivity}
\begin{figure}[!t]
    \centering
    \includegraphics[width=0.9\linewidth]{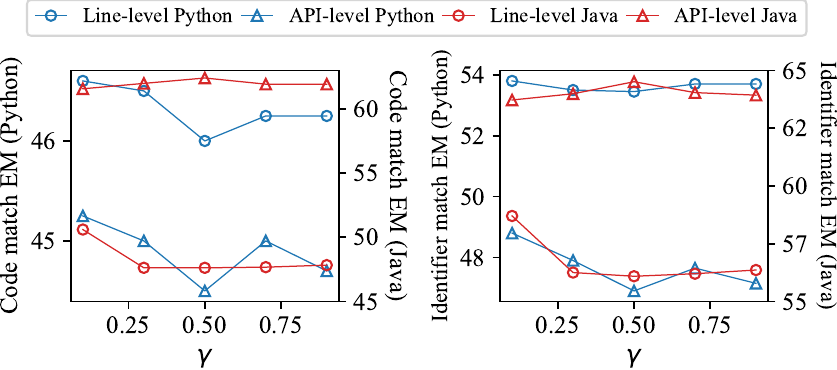}
    \vspace{-4mm}
    \caption{Hyper-parameter sensitivity.}
    \vspace{-4mm}
    \label{fig:hyperparameter}
\end{figure}

To conduct a more granular analysis of the impact of fine-grained step on the performance of \graphcoder, we demonstrate its performance as the hyperparameter $\gamma$ varies. $\gamma$ is the dependence distance shrink factor in the fine-grained step. 
A lower $\gamma$ places more emphasis on the local structure of the completion target. 
From Fig.~\ref{fig:hyperparameter}, we can observe that the performance of \graphcoder is robust to the hyper-parameter $\gamma$.
Specifically, the variation of code match EM when $\gamma$ changes from 0.1 to 0.9 is 0.60, 0.75, 2.98, and 0.84 on line-level, API-level Python, and line-level, API-level Java tasks, respectively.
Generally, the optimal $\gamma$ that yields the best performance depends on the intrinsic feature of tasks.
For example, on API-level Python tasks, the best $\gamma$ is 0.1, while that value on the API-level Java task is 0.5.
Although randomly selecting a $\gamma$ may lead to sub-optimal performance, \graphcoder is still likely superior to other baseline methods. 
By comparing Fig.~\ref{fig:hyperparameter} with Table~\ref{tab:completion-results}, \graphcoder's worst performance when varying $\gamma$
remains better than baseline methods.
 
\begin{tcolorbox}[left=2pt,right=2pt,top=0pt,bottom=0pt,boxrule=0.5pt]
\textbf{Answer to RQ3: }
\textit{
The performance of \graphcoder degrades after the removal of its CCG components or the coarse-to-fine steps. Among its various components, the CFG and the coarse-grained retrieval step are the most critical for its effectiveness.
}
\end{tcolorbox}

\subsection{RQ4: Cost}\label{sec:cost}
In this subsection, we compare the resource cost of \graphcoder with baseline RAG methods from three aspects: (1) time efficiency of retrieval; (2) database storage; and (3) number of tokens utilized.

\subsubsection{Time efficiency of retrieval}
To investigate the retrieval efficiency of \graphcoder and sliding window-based methods, we compare their end-to-end retrieval running time on an average of 8000 completion tasks in Table~\ref{tab:runtime-storage}.
Specifically, the running time comprises the time needed for converting code sequences into bag-of-words embedding (via a local tokenizer) and searching for the top-$k$ code snippets. 

Compared to sequence-based methods (Vanilla RAG, Shifted RAG, and RepoCoder), \graphcoder is more time-efficient. This is because the database of \graphcoder is statement-level, whereas that of sliding window-based methods is line-level. The statement-level database significantly reduces the number of entries by not storing blank lines or comments and by consolidating multi-line statements into single entries. This reduction in entries decreases the number of calculations required.
For RepoCoder, it is more time-consuming since it requires three iterations, each of which includes a sliding window-based search.
Additionally, since the fine-grained step is used for re-ranking, \graphcoder only needs to calculate the similarity between the query CCG and a small subset of entries in the database. 
This significantly reduces the running time for the fine-grained step. As shown in Table~\ref{tab:runtime-storage}, the fine-grained step accounts for only 3.06\% of the total retrieval time. 
To further examine the relationship between repository size and retrieval time, we present the running time for retrieval against the number of code lines in the repository in Fig.~\ref{fig:cost}. 
It can be observed from Fig.~\ref{fig:cost} that the retrieval time of \graphcoder increases more slowly compared to sliding window-based methods.

\subsubsection{Database storage}
\begin{figure}
    \centering
    \includegraphics[width=0.8\linewidth]{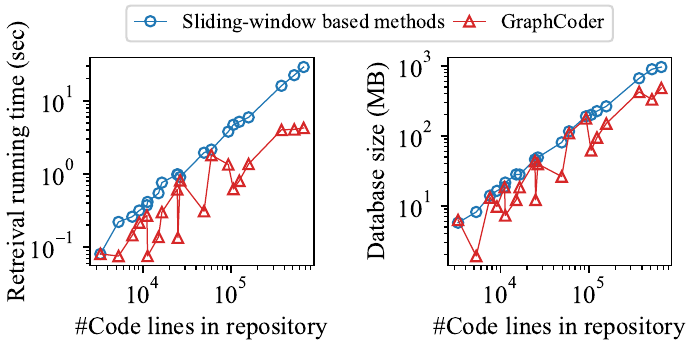}
    \vspace{-4mm}
    \caption{Running time and database size of sliding window based retrieval methods and \graphcoder.}
    \vspace{-2mm}
    \label{fig:cost}
\end{figure}

Table~\ref{tab:runtime-storage} demonstrates the database size of various RAG methods.
Since the Vanilla RAG, Shifted RAG, and RepoCoder are all sliding window-based methods with the same window size and stride, their database are of the same size.
As seen in Table~\ref{tab:runtime-storage}, \graphcoder reduces the number of entries by 79.5\% compared to sliding window-based methods by using a statement-level database instead of a line-level one.
Therefore, despite the need to store graph structures, \graphcoder's database is more space-saving.
Additionally, Fig.~\ref{fig:cost} exhibits the relationship between the number of code lines in the repository and the database size. For sliding window-based methods, the database size increases linearly with the number of lines, whereas \graphcoder's database size remains smaller than sliding window-based ones.

{\small
\begin{table}[]
    \centering
    \caption{Details of time efficiency and database storage.}
    \vspace{-4mm}
    \scalebox{0.8}{
    \begin{threeparttable}
    \begin{tabular}{lcccc}
    \toprule
         &  \multicolumn{1}{c}{Retrieval time} & \multicolumn{3}{c}{Database storage}\\
         & (sec) & \#Entries & Size (MB) & Prop.\\
    \hline
    Vanilla RAG & 4.7290 & 115109 & 159.4 & 9.7 \\
    Shifted RAG & 4.4894  & 115109 & 159.4 & 9.7 \\
    RepoCoder & 14.0168 & 115109 & 159.4 & 9.7 \\
    \graphcoder & 1.0753 & 23560 & 102.4 & 6.7 \\
    \  - Coarse & 1.0424 & - & - & - \\
    \  - Fine & 0.0329  & - & - & - \\
    \bottomrule
    \end{tabular}
            \begin{tablenotes}
        \footnotesize
        \vspace{-0.5mm}
        \item The results of RepoCoder are obtained after three iterations.
      \end{tablenotes}
    \end{threeparttable}}
    \label{tab:runtime-storage}
\end{table}
}

\subsubsection{Number of tokens}
To study the consumption for generation, we compare the number of tokens utilized of \graphcoder with other methods in Table~\ref{tab:token-consumption}.
It can be observed that, on average, the computational overhead of GraphCoder, in terms of input/output tokens, is lower than that of the other three retrieval-augmented methods: Vanilla RAG, Shifted RAG, and RepoCoder. However, the difference in the number of tokens is not significant, as we employ the same prompt template and organization method for retrieved results in the prompt.

{\small
\begin{table}[!tbp]
    \centering
    \caption{Number of input (\#In) and output (\#Out) tokens used.}
    \vspace{-4mm}
    \scalebox{0.8}{
    \begin{threeparttable}
    \begin{tabular}{lcccccc}
    \toprule
         &  \multicolumn{2}{c}{GPT3.5} & \multicolumn{2}{c}{StarCoder-15B} & \multicolumn{2}{c}{CodeGen-16B}\\
         & \#In & \#Out & \#In & \#Out & \#In & \#Out \\
         \hline
    No RAG & 758.18 & 93.87 & 803.57 & 55.87 & 701.12 & 70.99 \\
    Vanilla RAG	& 2557.12 & 67.01 & 3882.12 & 77.72 & 2239.95 & 64.64 \\
    Shifted RAG & 2990.92 & 87.99 & 3716.51 & 58.94 & 2311.48 & 63.71 \\
    RepoCoder & 7772.52 & 207.93 & 11500.86 & 234.06 & 6711.87 & 195.25 \\
    \graphcoder & 2666.13 & 88.14 & 3715.51 & 77.04 & 2201.08 & 71.29 \\
    \bottomrule
    \end{tabular}
        \begin{tablenotes}
        \footnotesize
        \vspace{-0.5mm}
        \item The results of RepoCoder are obtained after three iterations.
      \end{tablenotes}
    \end{threeparttable}}
    \label{tab:token-consumption}
\end{table}
}

\begin{tcolorbox}[left=2pt,right=2pt,top=0pt,bottom=0pt,boxrule=0.5pt]
\textbf{Answer to RQ4: }
\textit{
Compared to sliding window-based RAG methods, \graphcoder does not consume more tokens but is more efficient in terms of time and space by virtue of its statement-level database instead of a line-level one.
}
\end{tcolorbox}

\section{Threats to Validity}

\emph{Internal validity.}
The internal threats to validity lie in the implementation of baseline methods and the selection of code LLMs used in experiments.
For the implementation of baseline methods, we directly utilize source code from GitHub provided by RepoCoder~\cite{zhang2023repocoder} and configure it according to their paper to ensure a fair comparison.
As for other baseline methods (Vanilla RAG, Shifted RAG), we implement them by ourselves as they have no publicly available implementations. 
Considering that Vanilla RAG and Shifted RAG are also sliding window methods, similar to RepoCoder, we adopt RepoCoder’s implementation to mitigate internal threats brought by potential bugs.
For the selection of code LLMs, we are keen to use more recent newly released code LLMs, such as CodeLlama~\cite{roziere2023code} and DeepSeek-Coder~\cite{deepseek-coder}, to verify the effectiveness of \graphcoder.
Regrettably, these recent models pose a data leakage risk to  RepoEval~\cite{zhang2023repocoder} and even our newly constructed RepoEval-Updated.
Therefore, we meticulously select six suitable code LLMs without data leakage risk to validate \graphcoder's effectiveness, including OpenAI's GPT-3.5, StarCoder $15$B, CodeGen2 $1$B, $3.7$B, $7$B and $16$B, for the fairness of experiments.

\emph{External validity.}
The threat to external validity mainly lies in the generalizability of our method, including its ability to be applied to different programming languages and diverse repositories.
Our evaluations focus on open-source repositories in two mainstream programming languages: Python and Java, so our results are limited in this scope. 
For generalizability to other programming languages, \graphcoder can be migrated with minimal effort by first constructing the code context graph based on the abstract syntax tree produced by tree-sitter, and then directly using \graphcoder to achieve RAG-based code completion.
For the generalization to diverse repositories, we try our best to cover a wide range of repositories of different sizes.
However, there may exist potential threats to \graphcoder when the downstream evaluation repositories contain relatively low code duplication.
This is primarily because low duplication will significantly reduce the recall rate during the retrieval phase of \graphcoder.
To clearly delineate the performance boundaries, we offer a more detailed analysis on the impact of code duplication for \graphcoder's efficacy in our experiments to demonstrate its impact.

\section{Conclusion}
In this paper, we propose \graphcoder, a graph-based code completion framework for repository-level tasks.
\graphcoder uses a code context graph (CCG) to capture the completion target's relevant context.
The CCG is a statement-level multi-graph with control flow and data and control dependence edges. 
The retrieval is done through coarse-to-fine steps, involving filtering candidate code snippets and re-ranking them using a decay-with-distance structural similarity measure. 
After that, \graphcoder employs pre-trained language models to generate the next lines based on the retrieved snippets.
To comprehensively evaluate the performance of \graphcoder, we conduct experiments using 5 LLMs and 8000 code completion tasks sourced from 20 repositories.
Experimental results demonstrate \graphcoder's effectiveness, significantly improving the accuracy of code completion via more exactly matched code for generation with less retrieval time and space overhead.

\bibliographystyle{ACM-Reference-Format}

\end{document}